# A CMOS-based Tactile Sensor for Continuous Blood Pressure Monitoring

K.-U. Kirstein[1], J. Sedivy[2], T. Salo[1], C. Hagleitner[3], T. Vancura[1], A. Hierlemann[1]

[1]: Physical Electronics Laboratory, ETH Zurich, HPT-H4.1, CH-8093 Zurich, Switzerland
E-mail: kirstein@iqe.phys.ethz.ch

[2]: now at Czech Technical University, Faculty of Electrical Engineering, Dept. of Measurement, Prague, Czech Republic

[3]: now at IBM Research GmbH, Zurich Research Laboratory, CH-8803 Rüschlikon, Switzerland

## Abstract

*A monolithic integrated tactile sensor array is presented, which is used to perform non-invasive blood pressure monitoring of a patient. The advantage of this device compared to a hand cuff based approach is the capability of recording continuous blood pressure data. The capacitive, membrane-based sensor device is fabricated in an industrial CMOS-technology combined with post-CMOS micromachining. The capacitance change is detected by a ΣΔ-modulator. The modulator is operated at a sampling rate of 128kS/s and achieves a resolution of 12bit with an external decimation filter and an OSR of 128.*

## 1 Introduction

Current devices for blood pressure measurements show several drawbacks: External methods based on hand cuffs or other pressurized devices are only able to accomplish single measurements at a rate of some Hertz. Thus the continuous recording of a blood pressure waveform is not possible. Intravascular pressure sensors are capable of recording continuous blood pressure data, but they have to be implanted or need at least an open access to the vein of the patient. The presented sensor device enables a continuous recording of blood pressure extravascular or even percutaneously. It measures the displacement of a surface caused by the movement of a blood vessel wall, due to its overpressure inside, as shown Figure 1. This method, which is also known as tonometry, has been realized so far only with hybrid integration of sensor unit and signal conditioning circuitry, leading to very bulky solutions [1, 2]. The presented monolithic implementation of a blood pressure sensor overcomes this disadvantage and enables very compact and even portable devices. Additionally an invasive application, e.g., on the beating heart during surgery is also possible.

## 2 System Description

Figure 1 shows the application of the integrated sensor device for monitoring blood pressure of a vessel through the skin. The overpressure inside that blood vessel, vein or artery, causes a movement of the vessel wall. This movement leads to a local displacement of the skin surface. When a force sensor is applied at the right place of the surface, this displacement can be detected and gives a signal proportional to the blood pressure inside the underlying vessel.

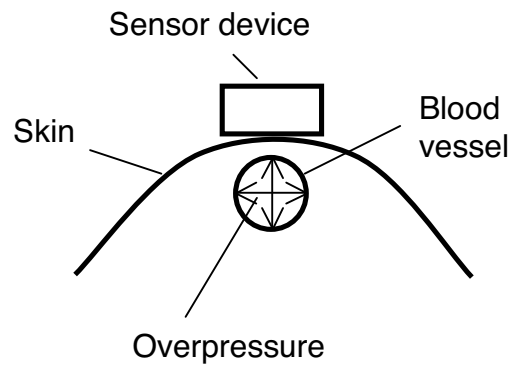

**Figure 1: Principle of monitoring blood pressure due to surface displacement**

In order to relax the necessary accuracy of sensor placement, an array of force detectors is used and the sensor element with the strongest signal is selected during measurement. This can also be used for localizing blood vessels, buried in tissue.

## 2.1 Sensor Array

On the chip there is an array of square-shaped force-sensitive elements. A single element has a suspended elastic membrane with a top electrode for a capacitive readout of deflection. The membrane is made of CMOS dielectric layers (silicon oxide/ nitride) and metallization (aluminum). The bottom electrode is of polysilicon (Figure 2). The side length and thickness of the membrane are 100$\mu$m and 3$\mu$m, respectively. The pitch of the membranes in the array is 150 $\mu$m.

In order to release the membranes, a potassium hydroxide (KOH) etch is applied from the back of the chip to create an opening in the silicon substrate. Through this opening it is possible to access the first metal layer of the CMOS process. This revealed metal layer is then selectively removed from the wafer stack. Thus the membranes are released and are free to deflect. The approach of releasing the membranes from the back of the wafer avoids post-release sealing present in similar sensor structures [3, 4]. The sealing is necessary to protect the devices from the hostile environment of bodily fluids [5].

For the application it is important to protect the electrical interconnections and provide a contacting surface to a biological tissue. This is carried out with a PDMS (polydimethylsiloxane) on top of the membrane array surrounded by gloptop epoxy to protect the electronics.

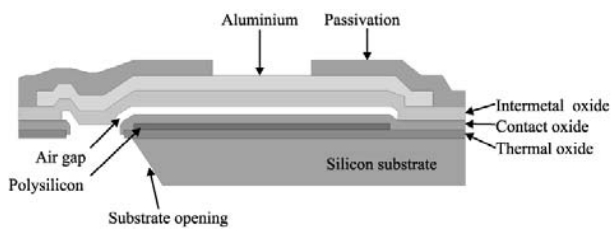

**Figure 2: Cross-section of the released membrane structure in standard CMOS-technology**

Force is applied to the top of the sensor element and this leads to a deformation of the membrane layers. This displacement is detected by the change of capacitance between the top electrode and the bottom electrode, formed by the second metal layer and the polysilicon layer of the CMOS-process, respectively.

## 2.2 Readout Circuit

The measured displacement signal is filtered and converted to a digital value by the on-chip readout electronics as shown in the block diagram of Figure 3.

The two-by-two sensor array is connected to a single bit, second order $\Sigma\Delta$-modulator via an analog multiplexer. This analog second order filter is realized by a fully-differential switched-capacitor filter. Figure 6 shows the two stage SC-filter and the connected sensor and reference capacitors. For sake of clarity, a single-ended implementation is shown. When a constant voltage is applied to the inputs of sensor and reference element, a signal, proportional to the difference of sensor and reference capacitor is integrated at the first stage of the $\Sigma\Delta$-modulator [6]. The output of the modulator is connected to an external digital decimation filter. Currently this filter is implemented in an FPGA, which also provides an interface (USB) to a computer system.

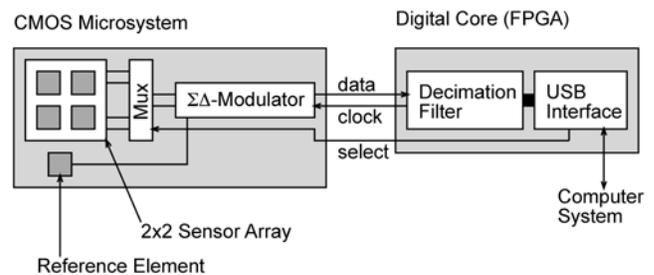

**Figure 3: Block diagram of the readout circuitry**

The transducer elements of a sensor array are connected via two synchronized analog multiplexers to the readout circuit as shown in Figure 4. This enables a modular design, which can be easily extended to larger array sizes. The settling when switching between different sensor elements is limited by the signal bandwidth of the $\Sigma\Delta$-AD-converter.

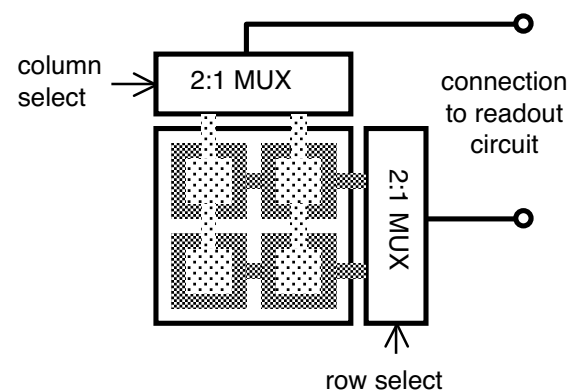

**Figure 4: 2x2 transducer array with multiplexer for selecting a sensor element**

## 3   Results

Figure 5 shows a micrograph of a demonstrator, fabricated in a standard 0.8 CMOS-technology with additional post-processing. The chip size is 2.6 x 1.9 mm$^2$. It incorporates a 2x2 array of the pressure transducers, a reference structure and the ΣΔ-modulator to readout the change of capacity. The ΣΔ-modulator additionally has a differential voltage interface, so a full characterization of the analog to digital conversion of this circuit can be accomplished, independent of the connected transducer.

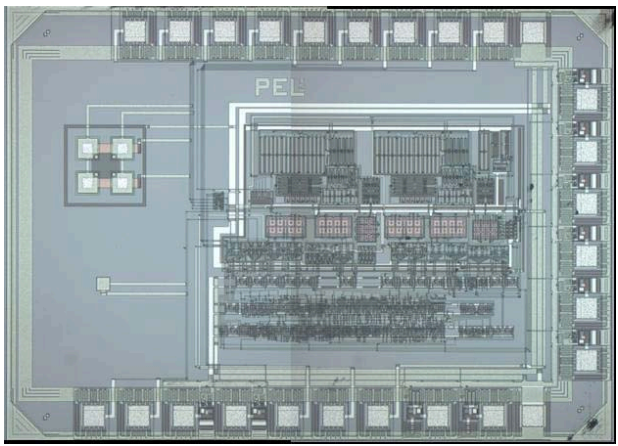

**Figure 5: Chip micrograph of the sensor device.**

The sensor chip is connected to an FPGA, which contains the digital decimation filter and an USB interface to deliver the measurement data to a computer system (Figure 3).

### 3.1   Electrical Measurements

To test the performance of the ΣΔ-analog-digital converter the additional voltage input is used. Figure 7 shows the spectrum of a converted sine-wave input signal.

The modulator was operated at a frequency of 128kHz and an oversampling ratio of 128 leading to a conversion rate of 1kS/s. The decimation filter was implemented as a two stage filter architecture, comprising a 3$^{rd}$ order SINC-filter as first stage and a 32 tap FIR-filter as second stage. The cutoff frequency of the filter is 500Hz and the output resolution is 12bit. As can be seen from Figure 6, a signal-to-noise ratio better than 72dB was achieved.

The power consumption of the sensor chip is 11.5mW at 5V supply voltage for 128kHz sampling frequency.

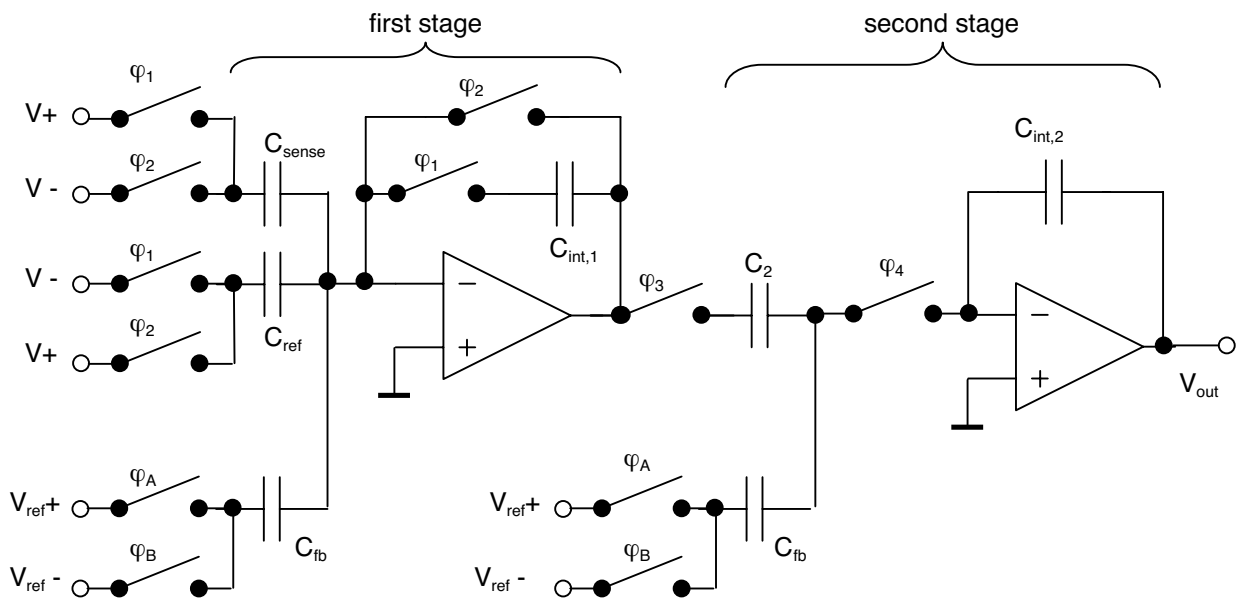

**Figure 6: Block diagram of the two stage ΣΔ-modulator with sensor and reference capacitors connected to the first stage**

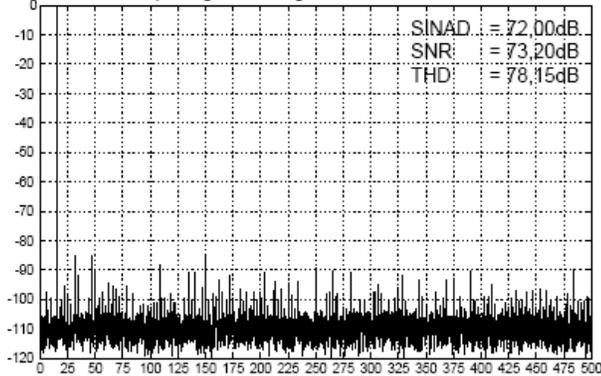

**Figure 7: Measured spectrum of the 12b ΣΔ-ADC at 15.625Hz**

### 3.2 Blood Pressure Measurement

Figure 8 shows the chip assembled on a PCB for blood pressure measurements. To apply a backpressure to the sensor membranes a pressure tube is connected to the backside of the sensor die. An applied overpressure bends the membrane layers upwards, so that they stick out and touch the surface of the measured object.

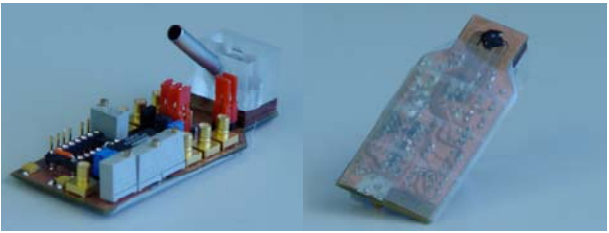

**Figure 8: Top and bottom view of the PCB with assembled sensor chip**

In Figure 9 a recorded blood pressure waveform is shown. The sensor device has been attached to a test person's wrist and the acquired signal is relative to the pressure applied to the skin surface by the sensor device. In order to get absolute pressure values, a calibration has to be performed. This calibration can be accomplished by measuring the systolic and diastolic pressure with a conventional hand cuff device (see Figure 9).

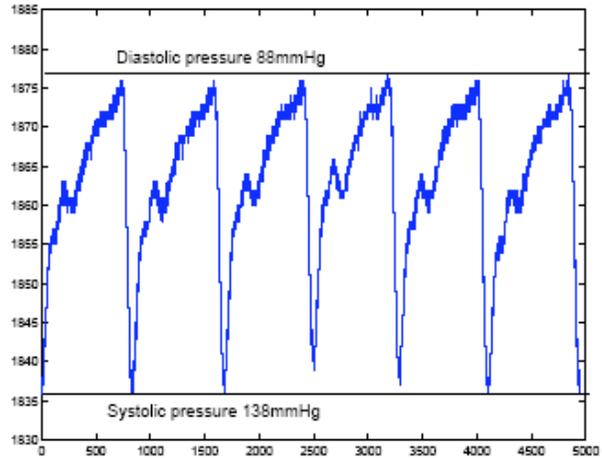

**Figure 9: Measured blood pressure wave with calibration of systolic and diastolic values**

## 4 Conclusion and Outlook

A novel monolithic pressure sensor for external blood pressure monitoring has been presented. The device is fabricated in a standard CMOS technology with additional post-processing steps. The integrated readout circuitry achieves a resolution of 12bit at a conversion rate of 1kS/s. Continuous blood pressure measurements using a CMOS-based tactile sensor array have been successfully demonstrated.

Future work will include an improvement of the resolution during blood pressure measurements. This can be achieved by adjusting the feedback capacitors of the first modulator stage. Also an increased conversion rate would be desirable. Field test have to be performed in order evaluate reliability and stability of blood pressure monitoring.

## 5 Acknowledgements

The authors thank Professor Henry Baltes (on leave) for sharing laboratory resources and for his ongoing stimulating interest in their work. Special thanks goes also to Dr. J. Grünenfelder from the University Hospital Zurich, Switzerland, for many fruitful discussions. This work is funded by the National Science Foundation, Switzerland and is part of the NSF-CoMe project 11, "Robotics in Cardiovascular Surgery".

## 6   References

[1] J. G. Webster (editor), *The Measurement, Instrumentation and Sensors Handbook*, CRC Press, 1999.

[2] L. A. Steiner, A. J. Johnston, R. Salvador, M. Czosnyka, D. K. Menon, "Validation of a tonometric noninvasive arterial blood pressure monitor in the intensive care setting", *Anaesthesia*, 58, pp. 448-454, 2003.

[3] B.L. Gray, R.S. Fearing, "A surface micromachined microtactile sensor array", *IEEE Int. Conf. Robotics and Automation*, 1996

[4] B.J. Kane, M.R. Cutkosky, G.T.A. Kovacs, "A traction stress sensor array for use in high-resolution robotic tactile imaging", *J. Microelectromech. Syst*. 9(4), pp. 425-434, 2000

[5] T. Salo, T. Vancura, O. Brand, H. Baltes, "CMOS-based Sealed Membranes for Medical Tactile Sensor Arrays", *Technical Digest MEMS 2003*, pp. 590-593.

[6] C. Hagleitner, D. Lange, A. Hierlemann, O. Brand, H. Baltes, „CMOS Single-chip gas detection system comprising capacitive, calorimetric and mass-sensitive microsensors", *IEEE Journal of Solid-State Circuits*, vol. 37, pp. 1867-1878, December 2002.